\documentclass[12pt]{article}

\catcode`\@=11
\@addtoreset{equation}{section}

\global\arraycolsep=2pt
\usepackage{mathrsfs,amsbsy,amssymb,latexsym,amsfonts,amsmath,cite}

\allowdisplaybreaks

\newcommand\CQ{{\mathcal Q}}

\newcommand\CI{{\mathcal I}}
\newcommand\CJ{{\mathcal J}}

\newcommand\diag{\mathrm{diag}}

\newcommand{\adss}[2]{{AdS$_{#1}\times$S$^{#2}$}}

\newcommand{\s}[1]{{S$^{#1}$}}

\newcommand\nn{\nonumber}
\newcommand\CN{{\mathcal N}}

\newcommand{\sh}{Schr\"odinger }

\begin{document}

\begin{flushright}
\parbox{4.2cm}
{OIQP-08-04 \hfill \\ 
NSF-KITP-08-69}
\end{flushright}

\vspace*{0.5cm}

\begin{center}
{\Large \bf 
Super Schr\"odinger algebra in AdS/CFT
}
\end{center}
\vspace{10mm}

\centerline{\large Makoto Sakaguchi$^{a}$
 and Kentaroh Yoshida$^{b}$
}

\vspace{8mm}

\begin{center}
$^a$ {\it Okayama Institute for Quantum Physics \\
1-9-1 Kyoyama, Okayama 700-0015, Japan} \\
{\tt makoto\_sakaguchi\_at\_pref.okayama.jp}
\vspace{5mm}

$^b$ {\it Kavli Institute for Theoretical Physics, \\ 
University of California, Santa Barbara,  \\
Santa Barbara CA.\ 93106, USA} 
\\
{\tt kyoshida\_at\_kitp.ucsb.edu}
\end{center}

\vspace{1cm}

\begin{abstract} 
We discuss (extended) super Schr\"odinger algebras
obtained as
subalgebras
of the superconformal algebra psu(2,2$|$4). The Schr\"odinger algebra
with two spatial dimensions can be embedded into so(4,2). In the
superconformal case the embedded algebra may be enhanced to the
so-called super Schr\"odinger algebra. In fact, we find an extended
super Schr\"odinger subalgebra of psu(2,2$|$4). It contains 
24 supercharges (i.e.,
3/4 of the original supersymmetries) and the generators of so(6), as well as
the generators of the original Schr\"odinger algebra.
 In
particular, the 24 supercharges come from 16 rigid supersymmetries and
half of 16 superconformal ones. Moreover, this superalgebra contains a
smaller super Schr\"odinger subalgebra, which is a supersymmetric extension of
the original Schr\"odinger algebra and so(6)
by eight supercharges (half of 16 rigid
supersymmetries).
It is still a subalgebra
even if there are no
so(6) generators.
We also discuss super Schr\"odinger subalgebras of the superconformal
algebras, osp(8$|$4) and osp(8$^{\ast}|$4).
\end{abstract}

\thispagestyle{empty}
\setcounter{page}{0}

\newpage

\section{Introduction}

AdS/CFT correspondence \cite{AdS/CFT,GKP,W} is increasingly
important from  fundamental aspects in string theory and its
applications to some realistic systems such as QCD, quark-gluon plasma
and condensed matter physics. The rigorous proof of the correspondence
has not been given yet, and it is still a conjecture. But it is firmly
supported by enormous evidence without any obvious failures. In recent
years the integrability of the AdS superstring \cite{BPR} has played an
important role in testing the AdS/CFT beyond supergravity
approximation. With this integrability and symmetry argument, the
S-matrix of the AdS superstring has now been proposed \cite{Beisert}.

\medskip 

It is, however, technically difficult to treat the full AdS superstring
\cite{MT}, and it is still a nice direction to look for a solvable limit
of the AdS superstring. The Penrose limit \cite{Penrose} is a well-known
example and  the resulting theory is a pp-wave
string \cite{Metsaev} which is exactly solvable \cite{MT2} with the
light-cone gauge fixing. The solvability was exploited to construct the
BMN operator correspondence \cite{BMN}. Another example is a
non-relativistic limit in the target spacetime \cite{GO}. 
In this
limit, the world-sheet theory of the AdS superstring becomes a set
of free theories on AdS$_2$ geometry. It is also exactly solvable
\cite{GGK,ST} and it has well been studied \cite{DGT,SY:NR,SY:NH}.

\medskip 

As a first step towards a new solvable limit of the AdS superstring, we
focus upon a \sh algebra \cite{Sch1,Sch2} obtained as a subalgebra of the
conformal algebra so(4,2). A \sh algebra is a non-relativistic analog of
the conformal algebra. When the \sh algebra has $d$ spatial dimensions,
it is embedded into a conformal algebra so($d$+2,2) as a subalgebra. In
this paper we discuss (extended) super \sh subalgebra of the
superconformal algebra psu(2,2$|$4). In the superconformal case the
embedded \sh algebra may be enhanced to a supersymmetric version of the
\sh algebra\footnote{
Although some supersymmetric extensions
of \sh algebra are discussed in \cite{GGT,Beckers,DH,SSch},
there would be no overlap with our result.
After this paper, less supersymmetric \sh algebras have been found in \cite{SY2}.
One of them coincides with the symmetry of \cite{LLM}.
}. It should be called (extended) super \sh algebra, where the
word ``extended'' implies the additional bosonic generators
coming from so(6) other than the original \sh.

\medskip 

As a new result, we find an extended super Schr\"odinger subalgebra of
psu(2,2$|4$). It contains 
24 supercharges (i.e., 3/4 of the original
supercharges) and those of so(6),
as well as the generators of the original
Schr\"odinger algebra. In particular, the 24 supercharges
come from 16 rigid supersymmetries and half of 16 superconformal
ones. This superalgebra further contains a smaller super Schr\"odinger
subalgebra, which is composed of generators of the original
Schr\"odinger algebra, eight supercharges (half of 16 rigid
supersymmetries) and those of so(6). It is still a subalgebra
even if there are no
so(6) generators. We also discuss super Schr\"odinger algebras in the
superconformal algebras related to the AdS$_{4/7}\times$S$^{7/4}$\,,
namely
osp(8$|$4) and osp(8$^{\ast}|$4). The results are similar to the case of
psu(2,2$|$4).

\medskip 
Recently
the non-relativistic CFT \cite{Henkel,MSW,SW,NS} is discussed
in relation to fermions at unitarity. The gravity background preserving
the \sh symmetry has been found in \cite{Son,BM} and it is a candidate
of the gravity dual of cold atoms\footnote{
An application to the aging phenomena is also discussed in \cite{Minic}.}. 
It is given by the geometry which is
conformally equivalent to an asymptotically plane-wave background. It would be an
interesting issue to embed this background into string theory as a
proper background, and discuss
the spectrum
of the string
theory. 
Hoping that our
results may be useful in this direction,
we leave it as a future problem.

\medskip 

This paper is organized as follows. In section 2 we introduce the
superconformal algebra psu(2,2$|$4) and identify (extended) super \sh
subalgebras. In section 3 we discuss other superconformal algebras,
osp(8$|$4) and osp(8$^{\ast}|$4). Section 4 is devoted to a conclusion
and discussions.

\section{Super Schr\"odinger algebras in psu(2,2$|$4)}

We begin with the superconformal algebra psu(2,2$|$4)
and then find (extended) super \sh subalgebras of it.

\subsection{Decomposition of psu(2,2$|$4)}

The (anti-)commutation relation of the super-\adss{5}{5} algebra, psu(2,2$|$4),
is composed as follows. The bosonic part contains the so(4,2) 
\begin{eqnarray}
&& [P_a,P_b]=J_{ab}~,~~~
[J_{ab},P_{c}]=\eta_{bc}P_{a}-\eta_{ac}P_{b}~, \nn \\ 
&& [J_{ab},J_{cd}]=\eta_{bc}J_{ad}+\mbox{3-terms} 
\qquad \qquad (a=0,1,2,3,4)\,, 
\end{eqnarray}
and the so(6)
\begin{eqnarray}
&& [P_{a'},P_{b'}]=-J_{a'b'}~,~~~
[J_{a'b'},P_{c'}]=\delta_{b'c'}P_{a'}-\delta_{a'c'}P_{b'}~, \nn\\ 
&& [J_{a'b'},J_{c'd'}]= \delta_{b'c'}J_{a'd'}+\mbox{3-terms} 
\qquad \qquad (a'=5,6,7,8,9)\,,
\label{s(6)} 
\end{eqnarray}
where $\eta_{ab}=\diag(-1,+1,+1,+1,+1)$
and $\delta_{a'b'}=\diag(+1,+1,+1,+1,+1)$.
Then the fermionic generator $\CQ$ satisfies
\begin{eqnarray}
&&
[P_a,\CQ]=-\frac{1}{2}\CQ\Gamma_a\CI i\sigma_2~,~~~
[P_{a'},\CQ]=\frac{1}{2}\CQ\Gamma_{a'}\CJ i\sigma_2~,~~~
[J_{AB},\CQ]=\frac{1}{2}\CQ\Gamma_{AB}~,\cr
&&
\{\CQ^T,\CQ\}=2iC\Gamma^Ah_+P_A
+iC\Gamma^{ab}\CI i\sigma_2 h_+ J_{ab} 
-iC\Gamma^{a'b'}\CJ i\sigma_2 h_+ J_{a'b'} 
\,,
\end{eqnarray}
where $A=(a,a')$ and 
\begin{eqnarray}
\CI \equiv\Gamma^{01234}\,, \quad \CJ \equiv \Gamma^{56789}\,. \nn
\end{eqnarray}
Here $\Gamma^A$'s are $(9+1)$-dimensional gamma-matrices
and $C$ is the charge conjugation matrix satisfying 
\[
\Gamma_A^T=-C\Gamma_AC^{-1}\,.
\]
The fermionic generator $\CQ$ is a pair of Majorana-Weyl spinors in
$(9+1)$-dimensions with the same chirality. 
$h_+$ is the chirality projector defined by
$h_+=\frac{1}{2}(1+\Gamma_{01\cdots 9})$\,.

\medskip

Let us define
\begin{eqnarray}
&&
\tilde P_\mu=\frac{1}{2}(P_\mu-J_{\mu4})\,, \quad 
\tilde K_\mu=\frac{1}{2}(P_\mu+J_{\mu4})\,,~~\tilde D=P_4\,,  \nonumber \\
&& \tilde J_{\mu\nu}=J_{\mu\nu}~,~~
\tilde Q=\CQ p_-\,, \quad 
\tilde S=\CQ p_+\,,
\end{eqnarray}
where $a=(\mu,4)$ and $\mu=0,1,2,3$.
Here the  projectors $p_\pm$ are
\begin{eqnarray}
p_\pm \equiv \frac{1}{2}(1\pm \Gamma^4\CI i\sigma_2)
=\frac{1}{2}(1\pm \Gamma^{0123}i\sigma_2)\,. 
\end{eqnarray}
Note that
\begin{eqnarray*}
p_\pm^TC=Cp_\mp~,~~~
\Gamma^{0123}i\sigma_2 p_\pm=\pm p_\pm~,~~~
[p_\pm,h_+]=0~.
\end{eqnarray*}
Then the (anti-) commutation relations are\footnote{
We suppress trivial commutators.}
\begin{eqnarray}
&&
[\tilde P_\mu,\tilde D]=-\tilde P_\mu\,, \quad 
[\tilde K_\mu,\tilde D]=\tilde K_\mu\,,  \quad 
[\tilde P_\mu, \tilde K_\nu]=\frac{1}{2}\tilde J_{\mu\nu}
+\frac{1}{2}\eta_{\mu\nu} \tilde D\,, \cr
&&
[\tilde J_{\mu\nu},\tilde P_\rho]
 =\eta_{\nu\rho}\tilde P_\mu-\eta_{\mu\rho}\tilde P_\nu\,, \quad 
[\tilde J_{\mu\nu},\tilde K_\rho]
 =\eta_{\nu\rho}\tilde K_\mu-\eta_{\mu\rho}\tilde K_\nu\,, \nn \\ 
&& 
[\tilde J_{\mu\nu}, \tilde J_{\rho\sigma}]
=\eta_{\nu\rho}\tilde J_{\mu\sigma}+\mbox{3-terms}\,,
\cr&&
\{\tilde Q^T,\tilde Q\}
=4iC\Gamma^\mu  p_-h_+\tilde P_\mu\,,  \quad 
\{\tilde S,\tilde S\}
=4iC\Gamma^\mu  p_+h_+\tilde K_\mu\,,  \cr
&&
\{\tilde Q^T,\tilde S\}
=iC\Gamma^{\mu\nu}\CI i\sigma_2p_+ h_+\tilde J_{\mu\nu}
+2i C\Gamma^4p_+h_+ \tilde D  \nn \\ 
&& \hspace*{2cm} +2iC\Gamma^{a'}p_+h_+ P_{a'}
-iC\Gamma^{a'b'}\CJ i\sigma_2 p_+h_+ J_{a'b'}\,, \cr 
&&
[\tilde P_\mu,\tilde S]=-\frac{1}{2}\tilde Q\Gamma_{\mu 4}\,, \quad 
[\tilde K_\mu,\tilde Q]=\frac{1}{2}\tilde S\Gamma_{\mu 4}\,, \quad 
[\tilde D, \tilde Q]=\frac{1}{2}\tilde Q\,,\quad 
[\tilde D, \tilde S]=-\frac{1}{2}\tilde S \,,
\cr
&&
[\tilde J_{\mu\nu}, \tilde Q]=\frac{1}{2}\tilde Q\Gamma_{\mu\nu}\,, \quad 
[\tilde J_{\mu\nu},\tilde S]=\frac{1}{2}\tilde S\Gamma_{\mu\nu}\,, \quad 
 \\&&
[P_{a'},\tilde Q]=\frac{1}{2}\tilde Q\Gamma_{a'}\CJ i\sigma_2\,, \quad 
[J_{a'b'},\tilde Q]=\frac{1}{2}\tilde Q\Gamma_{a'b'}\,, \quad 
\cr&&
[P_{a'},\tilde S]=\frac{1}{2}\tilde S\Gamma_{a'}\CJ i\sigma_2\,, \quad 
[J_{a'b'},\tilde S]=\frac{1}{2}\tilde S\Gamma_{a'b'}\,, \quad 
\label{so(6) fermionic}
\end{eqnarray}
and so(6) in \eqref{s(6)}. This is $\CN=4$ superconformal algebra in
four dimensions.  $\tilde Q$ are 16 supercharges while $\tilde S$
are 16 superconformal charges. 

\medskip

In order to clarify the embedding of the \sh algebra into the conformal
algebra, let us further decompose the generators as follows:
\begin{eqnarray}
&&
P_\pm=\frac{1}{\sqrt{2}}(\tilde P_0\pm \tilde P_3)~,~~
K_\pm=\frac{1}{\sqrt{2}}(\tilde K_0\pm \tilde K_3)~,~~
J_{i\pm}=\frac{1}{\sqrt{2}}(\tilde J_{i0}\pm \tilde J_{i3})~,~~
\cr&&
D=\frac{1}{2}(\tilde D-J_{03})~,~~
D'=\frac{1}{2}(\tilde D+J_{03})~,~~
P_i=\tilde P_i~,~~
K_i=\tilde K_i~,~~
J_{ij}=\tilde J_{ij}\,,
\end{eqnarray}
where $\mu=(0,i,3)$ with $i=1,2$.
The (anti-)commutation relations can straightforwardly be written down.
The bosonic part is
\begin{eqnarray}
&&
[J_{ij},J_{k\pm}]=\eta_{jk}J_{i\pm}-\eta_{ik}J_{j\pm}~,~~~
[J_{ij},P_k]=\eta_{jk}P_i-\eta_{ik}P_j~,~~~
[J_{ij},K_k]=\eta_{jk}K_i-\eta_{ik}K_j~,~~~ \nn \\ 
&& [J_{i\pm},J_{j\mp}]=J_{ij}\pm\eta_{ij}(D'-D)~,~~~
\cr&&
[P_i,K_j]=\frac{1}{2}J_{ij}+\frac{1}{2}\eta_{ij}(D'+D)~,~~~
[P_i,K_\pm]=\frac{1}{2}J_{i\pm}~,~~~
[P_\pm, K_i]=-\frac{1}{2}J_{i\pm}~,~~~
\cr&&
[D,J_{i\pm}]=\mp\frac{1}{2}J_{i\pm}~,~~~
[D',J_{i\pm}]=\pm\frac{1}{2}J_{i\pm}~,~~~
\cr&&
[P_i,J_{j\pm}]=\eta_{ij} P_\pm~,~~~
[K_i,J_{j\pm}]=\eta_{ij} K_\pm~,~~~
[J_{i\pm},P_\mp]=-P_i~,~~~
[J_{i\pm},K_\mp]=-K_i~,~~~
\cr&&
[P_+,K_-]=-D'~,~~~
[P_-,K_+]=-D~,~~~
\cr&&
[D,P_-]=P_-~,~~~
[D,P_i]=\frac{1}{2}P_i~,~~~
[D,K_+]=-K_+~,~~~
[D,K_i]=-\frac{1}{2}K_i~,~~~
\cr&&
[D',P_+]=P_+~,~~~
[D',P_i]=\frac{1}{2}P_i~,~~~
[D',K_-]=-K_-~,~~~
[D',K_i]=-\frac{1}{2}K_i~,
\end{eqnarray}
and so(6) in \eqref{s(6)}. The (anti-)commutation relations including
the fermionic generators are
\begin{eqnarray}
&&\{\tilde Q^T,\tilde Q\}=
4iC\Gamma^+p_-h_+P_+
+4iC\Gamma^-p_-h_+P_-
+4iC\Gamma^ip_-h_+P_i~,
\cr&&
\{\tilde S^T,\tilde S\}=
4iC\Gamma^+p_+h_+K_+
+4iC\Gamma^-p_+h_+K_-
+4iC\Gamma^ip_+h_+K_i~,
\cr&&
\{\tilde Q^T,\tilde S\}=
iC\Gamma^{ij}\CI i\sigma_2 p_+h_+ J_{ij}
+2iC\Gamma^{i+}\CI i\sigma_2 p_+h_+ J_{i+}
+2iC\Gamma^{i-}\CI i\sigma_2 p_+h_+ J_{i-}
\cr&&~~~~~~~~~~~~
-2iC\Gamma^4\Gamma^{+}\Gamma^-p_+h_+D'
-2iC\Gamma^4\Gamma^{-}\Gamma^+ p_+h_+D
\cr&&~~~~~~~~~~~~
+2iC\Gamma^{a'}p_+h_+ P_{a'}
-iC\Gamma^{a'b'}\CJ i\sigma_2 p_+h_+ J_{a'b'}
~,
\cr&& \hspace*{-0.3cm}
[K_\pm,\tilde Q]=-\frac{1}{2}\tilde S\Gamma^{\mp}\Gamma_4~,~~~
[K_i,\tilde Q]=\frac{1}{2}\tilde S \Gamma_{i4} ~,~~~
[P_\pm,\tilde S]=\frac{1}{2}\tilde Q\Gamma^{\mp}\Gamma_4~,~~~
[P_i,\tilde S]=-\frac{1}{2}\tilde Q \Gamma_{i4} ~,~~~
\cr&& \hspace*{-0.3cm}
[J_{ij},\tilde Q]=\frac{1}{2}\tilde Q \Gamma_{ij} ~,~~~
[J_{ij},\tilde S]=\frac{1}{2}\tilde S \Gamma_{ij} ~,~~~
[J_{i\pm},\tilde Q]=-\frac{1}{2}\tilde Q \Gamma_i\Gamma^\mp~,~~~
[J_{i\pm},\tilde S]=-\frac{1}{2}\tilde S \Gamma_i \Gamma^\mp~,
\cr&& \hspace*{-0.3cm}
[D,\tilde Q]=-\frac{1}{4}\tilde  Q \Gamma^{+}\Gamma^-~,~~
[D,\tilde S]=\frac{1}{4}\tilde  S \Gamma^-\Gamma^+~,~~
[D',\tilde Q]=-\frac{1}{4}\tilde  Q \Gamma^-\Gamma^+~,~~
[D',\tilde S]=\frac{1}{4}\tilde  S \Gamma^+\Gamma^-~, \nn 
\end{eqnarray}
and \eqref{so(6) fermionic}.  Here
$\Gamma^\pm \equiv
\frac{1}{\sqrt{2}}(\Gamma^0\pm\Gamma^3)$\,.

\subsection{Super Schr\"odinger subalgebras}

From
the superconformal algebra obtained in the
previous subsection, 
it is easy to see that the following subset of the
generators
\begin{eqnarray}
\{J_{ij},~J_{i+},~D, ~P_\pm,~P_i,~K_+\}
\label{Sch generators}
\end{eqnarray}
forms the Schr\"odinger algebra.
Its commutation relations are\footnote{
Denoting 
$(P_-,P_+,J_{i+},D,K_+)$
as
$(iH,iN,K_i,-\frac{i}{2}\hat D, \frac{i}{2}C)$,
we arrive at the expression used in the literature 
\begin{eqnarray*}
&&
[J_{ij},K_k]=-i\eta_{jk}K_i+i\eta_{ik}K_j~,~~~
[J_{ij},P_k]=-i\eta_{jk}P_i+i\eta_{ik}P_j~,~~~
\cr&&
[P_i,C]=-iK_i~,~~~
[P_i,K_j]=i\eta_{ij} N~,~~~
[K_i,H]=iP_i~,~~~
[H,C]=-i \hat D~,~~~
\cr&&
[\hat D,K_i]=-iK_i~,~~~
[\hat D,H]=2iH~,~~~
[\hat D,P_i]=iP_i~,~~~
[\hat D,C]=-2iC~.
\end{eqnarray*}
}
\begin{eqnarray}
&&
[J_{ij},J_{k+}]=\eta_{jk}J_{i+}-\eta_{ik}J_{j+}~,~~~
[J_{ij},P_k]=\eta_{jk}P_i-\eta_{ik}P_j~,~~~ 
[P_i,K_+]=\frac{1}{2}J_{i+}~,~~~ 
\cr
&& [P_i,J_{j+}]=\eta_{ij} P_+~,~~~
[J_{i+},P_-]=-P_i~,~~~
[P_-,K_+]=- D~,~~~
\cr&&
[D,J_{i+}]=-\frac{1}{2}J_{i+}~,~~~
[D,P_-]=P_-~,~~~
[D,P_i]=\frac{1}{2}P_i~,~~~
[D,K_+]=-K_+~. 
\label{Sch}
\end{eqnarray}
This subalgebra may be enhanced to a supersymmetric subalgebra of
psu(2,2$|$4), including the so(6) part. It should be called
(extended) super \sh subalgebra.

\medskip 

We will look for super \sh subalgebras of psu(2,2$|$4). We
begin with the anti-commutation relation $\{\tilde S^T,\tilde S\}$, which
contains $K_-$ and $K_i$ in the right-hand side.  $K_-$ and $K_i$ are not in
\eqref{Sch generators}.  For some part of $\tilde S$ to be supercharges
of a super Schr\"odinger algebra, 
it is necessary to 
introduce some projectors
which project out the terms including $K_-$ and $K_i$ in $\{\tilde S^T,\tilde S\}$.
An example is
the light-cone projectors
\begin{eqnarray}
\ell_\pm=\frac{1}{2}(1\pm \Gamma^{03})=
-\frac{1}{2}\Gamma^\pm\Gamma^{\mp}~,
\end{eqnarray}
which commute with $h_+$ and $p_\pm$.
Then $\tilde S$ can be decomposed as 
\begin{eqnarray}
\tilde S= S +S' ~,~~~
S=\tilde  S\ell_-~,~~S'=\tilde S\ell_+~,
\label{tilde S decom}
\end{eqnarray}
and then $\{S^T,S\}$ contains only $K_+$, namely 
$\{S^T,S\}\sim
K_+~ $.  One may expect that $S$ is a supercharge of a super
Schr\"odinger algebra. We will see below that this is the case.

\medskip 

The anti-commutation relations between
$\tilde Q$ and $S$  are 
\begin{eqnarray}
&&\{\tilde Q^T,\tilde Q\}=
4iC\Gamma^+p_-h_+P_+
+4iC\Gamma^-p_-h_+P_-
+4iC\Gamma^ip_-h_+P_i~,
\cr&&
\{S^T,S\}=4iC\Gamma^+\ell_-p_+h_+ K_+~,
\cr&&
\{\tilde Q^T, S\}=
iC\Gamma^{ij}\CI i\sigma_2 \ell_-p_+h_+ J_{ij}
+2iC\Gamma^{i+}\CI i\sigma_2 \ell_-p_+h_+ J_{i+}
-2iC\Gamma^4\Gamma^{-}\Gamma^+ \ell_-p_+h_+D
\cr&&~~~~~~~~~~~~
+2iC\Gamma^{a'}\ell_-p_+h_+ P_{a'}
-iC\Gamma^{a'b'}\CJ i\sigma_2\ell_- p_+h_+ J_{a'b'}
~, 
\label{QQ}
\end{eqnarray}
where $\Gamma^\pm\ell_\pm=0~$ has been used.
The so(6) generators appear in the right-hand side of $\{\tilde Q^T,S\}$
and hence those should be added to \eqref{Sch generators}.

\medskip 

Next the commutation relations between
\eqref{Sch generators}
including so(6) generators, and $(\tilde Q, S)$ are
\begin{eqnarray}
&&
[K_+,\tilde Q]=-\frac{1}{2}S\Gamma^{-}\Gamma_4~,~~~
[P_-,S]=\frac{1}{2}\tilde Q\ell_+\Gamma^{+}\Gamma_4~,~~~
[P_i, S]=-\frac{1}{2}\tilde Q\ell_- \Gamma_{i4} ~,~~~
\cr&&
[J_{ij},\tilde Q]=\frac{1}{2}\tilde Q \Gamma_{ij} ~,~~~
[J_{ij}, S]=\frac{1}{2} S \Gamma_{ij} ~,~~~
[J_{i+},\tilde Q]=-\frac{1}{2}\tilde Q \Gamma_i\Gamma^-~,~~~
\cr&&
[D,\tilde Q]=-\frac{1}{4}\tilde  Q \Gamma^{+}\Gamma^-~,~~~
[D,  S]=\frac{1}{4}   S \Gamma^-\Gamma^+~,
\cr&&
[P_{a'},\tilde Q]=\frac{1}{2}\tilde Q\Gamma_{a'}\CJ i\sigma_2~,~~~
[J_{a'b'},\tilde Q]=\frac{1}{2}\tilde Q\Gamma_{a'b'}~,~~~
\cr&&
[P_{a'}, S]=\frac{1}{2}  S\Gamma_{a'}\CJ i\sigma_2~,~~~
[J_{a'b'}, S]=\frac{1}{2} S\Gamma_{a'b'}~.
\label{BQ}
\end{eqnarray}
Now $\tilde Q$ and $S$ only are contained in the right-hand sides
of the commutation relations,
and hence 
the set of generators
\begin{eqnarray*}
\{J_{ij},~J_{i+},~D, ~P_\pm,~P_i,~K_+,~
P_{a'},~J_{a'b'},~
\tilde Q,~S~\}
\end{eqnarray*}
forms an extended super
Schr\"odinger algebra. The bosonic subalgebra is a direct sum of the
Schr\"odinger algebra and so(6). 
The word ``extended'' is attached due to
the presence of the extra so(6) in addition to the original
\sh algebra. The number of the supercharge is 24 since 
1/4 supercharges have been projected out.

\medskip 

Finally we shall comment on super subalgebras of the above extended super
Schr\"odinger algebra.
$\tilde Q$ is also decomposed
as 
\begin{eqnarray}
\tilde Q=Q  + Q'\,,~~~
Q=\tilde Q \ell_-~,~~
Q'=\tilde Q\ell_+\,.
\label{tilde Q decom}
\end{eqnarray} 
Substituting it into the commutation relations,
it is easy to find that
the set of generators,
\eqref{Sch generators}, so(6) generators and $Q$,
 forms a subalgebra of
the extended super Schr\"odinger algebra. The bosonic generators form
the Schr\"odinger algebra and so(6). The (anti-)commutation relations
including $Q$ are
\begin{eqnarray}
&&
\{Q^T,Q\}= 4iC\Gamma^+\ell_-p_-h_+ P_+\,, \quad 
[J_{ij},Q]=\frac{1}{2}Q\Gamma_{ij}\,, 
\label{sub QQ} \\ &&
[P_{a'},Q]=\frac{1}{2}Q\Gamma_{a'}\CJ i\sigma_2\,, \quad 
[J_{a'b'},Q]=\frac{1}{2}Q \Gamma_{a'b'}\,. 
\end{eqnarray}

\medskip 

It is possible to further reduce the algebra keeping only \eqref{Sch generators}
and $Q$. The (anti-)commutation relations are \eqref{Sch} and \eqref{sub
QQ}. This is a supersymmetric extension of the Schr\"odinger algebra with eight
supersymmetries,
 of which bosonic subalgebra is the Schr\"odinger algebra
only.

\subsection{Interpretation of so(6) in (2+1) dimensions}

The so(6) in (2.2) corresponds to the isometry of \s{5}.  The
supercharge $\CQ$ is a pair of 16 component Majorana-Weyl spinors in
$(9+1)$-dimensions.  Under the reduction (2.4), $\CQ$ is decomposed into
a pair of four  four-component spinors, supercharges $\tilde Q$
and superconformal charges $\tilde S$, in
$(3+1)$-dimensions.  As seen in (2.6), so(6) $\cong$ su(4) rotates four of
$\tilde Q$ and four of $\tilde S$, separately.  Thus so(6) is related to
the su(4) R-symmetry of the $\CN=4$ superconformal algebra in
$(3+1)$-dimensions.  By the reduction, (2.7), \eqref{tilde S decom} and
\eqref{tilde Q decom}, four of $\tilde Q$ reduce to
a pair of
four two-component spinors,
$Q$ and 
$Q'$, in
$(2+1)$-dimensions.  Similarly $\tilde S$ reduces to a pair of four
two-component spinors, $S$ and $S'$.  As can be seen from (2.14), so(6)
rotates $Q$, $Q'$, $S$ and $S'$ separately.  Thus even in the extended
super Schr\"odinger algebra, so(6) acts as su(4) R-symmetry. The
commutation relations in (2.17) mean that so(6) $\cong$ su(4) acts on
four of two-component supercharges in ($2+1$)-dimensions.

\section{Other cases - osp(8$|$4) and osp(8$^*|$4)}

We consider the super-\adss{q+2}{9-q} algebras with $q=2$
and $5$\,,
that is, osp(8$|$4) and osp(8$^*|$4). These are the super
isometries of the near-horizon geometries of M2-brane and M5-brane,
respectively. For osp(8$|$4) and osp(8$^*|$4), 
super \sh
subalgebras are found in the same way as psu(2,2$|$4).

\subsection{osp(8$|$4) and osp(8$^*|$4)}

Let us first introduce the superconformal algebras, osp(8$|$4) and
osp(8$^*|$4).

\medskip 

The (anti-)commutation relations of the super-\adss{4}{7} algebra, i.e.,
osp(8$|$4), are
\begin{eqnarray}
&&
[P_a,P_b]=4J_{ab}~,~~~
[J_{ab},P_{c}]=\eta_{bc}P_{a}-\eta_{ac}P_{b}~,~~~
[J_{ab},J_{cd}]=\eta_{bc}J_{ad}+\mbox{3-terms}~,
\label{so(2,3)}
\\&& 
[P_{a'},P_{b'}]=-J_{a'b'}~,~~~
[J_{a'b'},P_{c'}]=\delta_{b'c'}P_{a'}-\delta_{a'c'}P_{b'}~,
\nn \\ && 
[J_{a'b'},J_{c'd'}]=
\delta_{b'c'}J_{a'd'}+\mbox{3-terms}~, 
\label{s(8)}
\\&&
[P_a,\CQ]=-\CQ\CI \Gamma_a~,~~~
[P_{a'},\CQ]=-\frac{1}{2}\CQ\CI \Gamma_{a'}~,~~~
[J_{AB},\CQ]=\frac{1}{2}\CQ\Gamma_{AB}~,\cr
&&
\{\CQ^T,\CQ\}=-2C\Gamma^AP_A
+2C\CI\Gamma^{ab} J_{ab}
-C\CI\Gamma^{a'b'} J_{a'b'}~, \qquad \CI \equiv \Gamma^{0123}\,, 
\label{algebra Q q=2}
\end{eqnarray}
where $a=0,1,2,3$, $a'=4,\cdots,9,\natural$
and $A=(a,a')$\,. The
commutation relations in \eqref{so(2,3)} and \eqref{s(8)} are those of
so(3,2) and so(8), respectively.  The gamma matrices $\Gamma^A$'s are
$(10+1)$-dimensional ones and the charge conjugation matrix $C$
satisfies the relation $\Gamma_A^T=-C\Gamma_AC^{-1}$.  The fermionic
generator $\CQ$ is a 32 component Majorana spinor in $(10+1)$-dimensions.

\medskip 

On the other hand, the (anti-)commutation relations of the super-\adss{7}{4}
algebra, i.e., osp(8$^*|$4), are given by 
\begin{eqnarray}
&&
[P_a,P_b]=J_{ab}~,~~~
[J_{ab},P_{c}]=\eta_{bc}P_{a}-\eta_{ac}P_{b}~,~~~
[J_{ab},J_{cd}]=\eta_{bc}J_{ad}+\mbox{3-terms}~,
\label{so(2,6)}
\\&&
[P_{a'},P_{b'}]=-4J_{a'b'}~,~~~
[J_{a'b'},P_{c'}]=\delta_{b'c'}P_{a'}-\delta_{a'c'}P_{b'}~, \nn \\ && 
[J_{a'b'},J_{c'd'}]=
\delta_{b'c'}J_{a'd'}+\mbox{3-terms}~,~~~~~
\label{s(5)}
\\&&
[P_a,\CQ]=-\frac{1}{2}\CQ\CI \Gamma_a~,~~~
[P_{a'},\CQ]=-\CQ\CI \Gamma_{a'}~,~~~
[J_{AB},\CQ]=\frac{1}{2}\CQ\Gamma_{AB}~,\cr
&&
\{\CQ^T,\CQ\}=-2C\Gamma^AP_A
-C\CI\Gamma^{ab} J_{ab}
+2C\CI\Gamma^{a'b'} J_{a'b'}~,
\qquad
\CI\equiv \Gamma^{789\natural}~,
\label{algebra Q q=5}
\end{eqnarray}
where $a=0,\cdots,6$, $a'=7,8,9,\natural$ and $A=(a,a')$.
The commutation relations in
\eqref{so(2,6)} and \eqref{s(5)} are those of so(6,2) and so(5),
respectively.

\subsection{Bosonic part}

Let us scale generators as follows
\begin{eqnarray}
&&P_a \to 2 P_a~~~\mbox{for}~~q=2
~~~~~\mbox{and}~~~~~
P_{a'}\to 2 P_{a'}~~~\mbox{for}~~q=5~.
\label{scaling P}
\end{eqnarray}
Then the bosonic subalgebra takes the standard form
\begin{eqnarray}
&&
[P_a,P_b]=J_{ab}~,~~~
[J_{ab},P_{c}]=\eta_{bc}P_{a}-\eta_{ac}P_{b}~,~~~
[J_{ab},J_{cd}]=\eta_{bc}J_{ad}+\mbox{3-terms}~,
\label{AdS}
\\&&
[P_{a'},P_{b'}]=-J_{a'b'}~,~~~
[J_{a'b'},P_{c'}]=\delta_{b'c'}P_{a'}-\delta_{a'c'}P_{b'}~,~~~
\nn \\ && 
[J_{a'b'},J_{c'd'}]=
\delta_{b'c'}J_{a'd'}+\mbox{3-terms}~,~~~~
\label{S}
\end{eqnarray}
where $a=0,\cdots,q+1$ and $a'=q+2,\cdots,9,\natural$.
The commutation relation in \eqref{AdS}  is so($q+1$,2), 
while that of \eqref{S} is so($10-q$).

\medskip 

By decomposing the generators as
\begin{eqnarray}
&&
\tilde P_\mu=\frac{1}{2}(P_\mu-J_{\mu q+1})~,~~
\tilde K_\mu=\frac{1}{2}(P_\mu+J_{\mu q+1})~,~~
\tilde D=P_{q+1}~,~~
\tilde J_{\mu\nu}=J_{\mu\nu}~,~~
\label{decomposition 1} \\
&& \qquad \qquad a=(\mu,q+1)~~\mbox{with}~~ \mu=0,\cdots,q
\,,  \nn 
\end{eqnarray}
the commutation relations in \eqref{AdS} become 
\begin{eqnarray}
&&
[\tilde P_\mu,\tilde D]=-\tilde P_\mu~,~~~
[\tilde K_\mu,\tilde D]=\tilde K_\mu~,~~~
[\tilde P_\mu, \tilde K_\nu]=\frac{1}{2}\tilde J_{\mu\nu}
+\frac{1}{2}\eta_{\mu\nu} \tilde D~,~~~
\cr&&
[\tilde J_{\mu\nu},\tilde P_\rho]
 =\eta_{\nu\rho}\tilde P_\mu-\eta_{\mu\rho}\tilde P_\nu~,~~~
[\tilde J_{\mu\nu},\tilde K_\rho]
 =\eta_{\nu\rho}\tilde K_\mu-\eta_{\mu\rho}\tilde K_\nu~,~~~
\cr&&
[\tilde J_{\mu\nu}, \tilde J_{\rho\sigma}]
=\eta_{\nu\rho}\tilde J_{\mu\sigma}+\mbox{3-terms}~.~~~~~~
\label{conformal}
\end{eqnarray}
This is the conformal algebra in $d=q+1$ dimensions.  
Further decomposition
\footnote{
Note that $J_{ij}$ vanishes for $q=2$ since $i=1$\,. 
}
\begin{eqnarray}
&&
P_\pm=\frac{1}{\sqrt{2}}(\tilde P_0\pm \tilde P_q)\,, \quad 
K_\pm=\frac{1}{\sqrt{2}}(\tilde K_0\pm \tilde K_q)\,, \quad 
J_{i\pm}=\frac{1}{\sqrt{2}}(\tilde J_{i0}\pm \tilde J_{iq})\,, 
\cr&&
D=\frac{1}{2}(\tilde D-J_{0q})\,,~~~
D'=\frac{1}{2}(\tilde D+J_{0q})\,,~~~
P_i=\tilde P_i\,,~~~
K_i=\tilde K_i\,,~~~
J_{ij}=\tilde J_{ij}\,, \qquad 
\label{decomposition 2} \\
&& \qquad \qquad \mu=(0,i,q)~~\mbox{with}~~i=1,\cdots,q-1\,, \nn 
\end{eqnarray}
leads to
\begin{eqnarray}
&&
[J_{ij},J_{kl}]=\eta_{jk}J_{il}+\mbox{3-terms}~,~~~
[J_{ij},J_{k\pm}]=\eta_{jk}J_{i\pm}-\eta_{ik}J_{j\pm}~,~~~
\cr&& \hspace*{-0.2cm}
[J_{ij},P_k]=\eta_{jk}P_i-\eta_{ik}P_j~,~~
[J_{ij},K_k]=\eta_{jk}K_i-\eta_{ik}K_j~,~~
[J_{i\pm},J_{j\mp}]=J_{ij}\pm\eta_{ij}(D'-D)~,~~~
\cr&&
[P_i,K_j]=\frac{1}{2}J_{ij}+\frac{1}{2}\eta_{ij}(D'+D)~,~~~
[P_i,K_\pm]=\frac{1}{2}J_{i\pm}~,~~~
[P_\pm, K_i]=-\frac{1}{2}J_{i\pm}~,~~~
\cr&&
[D,J_{i\pm}]=\mp\frac{1}{2}J_{i\pm}~,~~~
[D',J_{i\pm}]=\pm\frac{1}{2}J_{i\pm}~,~~~
\cr&&
[P_i,J_{j\pm}]=\eta_{ij} P_\pm~,~~~
[K_i,J_{j\pm}]=\eta_{ij} K_\pm~,~~~
[J_{i\pm},P_\mp]=-P_i~,~~~
[J_{i\pm},K_\mp]=-K_i~,~~~
\cr&&
[P_+,K_-]=-D'~,~~~
[P_-,K_+]=-D~,~~~
\cr&&
[D,P_-]=P_-~,~~~
[D,P_i]=\frac{1}{2}P_i~,~~~
[D,K_+]=-K_+~,~~~
[D,K_i]=-\frac{1}{2}K_i~,~~~
\cr&&
[D',P_+]=P_+~,~~~
[D',P_i]=\frac{1}{2}P_i~,~~~
[D',K_-]=-K_-~,~~~
[D',K_i]=-\frac{1}{2}K_i~.
\end{eqnarray}

Here the following set of the generators 
\[
\{J_{ij},~J_{i+},~D, ~P_\pm,~P_i,~K_+\}
\] 
forms a subalgebra of so($q$+1,\,2), whose commutation relations are
\begin{eqnarray}
&& \hspace*{-0.2cm}
[J_{ij},J_{kl}]=\eta_{jk}J_{il}+\mbox{3-terms}~,~~
[J_{ij},J_{k+}]=\eta_{jk}J_{i+}-\eta_{ik}J_{j+}~,~~
[J_{ij},P_k]=\eta_{jk}P_i-\eta_{ik}P_j~,~~~
\cr&&
[P_i,K_+]=\frac{1}{2}J_{i+}~,~~~
[P_i,J_{j+}]=\eta_{ij} P_+~,~~~
[J_{i+},P_-]=-P_i~,~~~
[P_-,K_+]=- D~,~~~
\cr&&
[D,J_{i+}]=-\frac{1}{2}J_{i+}~,~~~
[D,P_-]=P_-~,~~~
[D,P_i]=\frac{1}{2}P_i~,~~~
[D,K_+]=-K_+~.
\label{Sch algebra}
\end{eqnarray}
This is nothing but the
Schr\"odinger algebra with $q-1$ spatial directions (see footnote 3).

\subsection{Fermionic part}

The remaining task is to consider the fermionic part. Here 
the cases with $q=2$ and $q=5$ are discussed at once. 
For simplicity, let us
rescale $\CQ$ as $\CQ\to\sqrt{2}\CQ$ for $q=2$ and take the rescaling \eqref{scaling P}
 for the
bosonic generators.

\medskip 

Then the (anti-)commutation relations including $\CQ$,
\eqref{algebra Q q=2} and \eqref{algebra Q q=5},
 are rewritten as
\begin{eqnarray}
&&
\{\CQ^T,\CQ\}=-2C\Gamma^aP_a
+\epsilon C\CI\Gamma^{ab}J_{ab}
-\beta C\Gamma^{a'}P_{a'}
-\epsilon\frac{\beta}{2}C\CI\Gamma^{a'b'}J_{a'b'}~,~~~
\cr&&
[P_a,\CQ]=-\frac{1}{2}\CQ\CI \Gamma_a~,~~~
[J_{ab},\CQ]=\frac{1}{2}\CQ\Gamma_{ab}~,~~~
\cr&&
[P_{a'},\CQ]=-\frac{1}{2}\CQ\CI \Gamma_{a'}~,~~~
[J_{a'b'},\CQ]=\frac{1}{2}\CQ\Gamma_{a'b'}~,
\label{QQM}
\end{eqnarray}
where $\epsilon$\,, $\beta$ and $\CI$ are
defined as, respectively,
\begin{eqnarray}
\epsilon=\left\{
  \begin{array}{cl}
    1   &~~\mbox{for}~~ q=2   \\
    -1   &~~\mbox{for}~~ q=5   \\
  \end{array}
\right.
,~~~
\beta=\left\{
  \begin{array}{cl}
    1   &~~\mbox{for}~~ q=2   \\
    4   &~~\mbox{for}~~ q=5   \\
  \end{array}
\right.
,~~~
\CI=\left\{
  \begin{array}{cl}
    \Gamma^{0123}   &~~\mbox{for}~~ q=2   \\
    \Gamma^{789\natural}   &~~\mbox{for}~~ q=5   \\
  \end{array}
\right.
.
\end{eqnarray} 
Under the decomposition \eqref{decomposition 1} and 
\begin{eqnarray}
\CQ=\tilde Q +\tilde S ~,~~~
\tilde Q=\CQ p_-~,~~~
\tilde S=\CQ p_+~,~~~
p_\pm=\left\{
  \begin{array}{ll}
    \frac{1}{2}(1\pm \Gamma^{012})   &~~\mbox{for}~~ q=2   \\
    \frac{1}{2}(1\pm \Gamma^{6789\natural})    &~~\mbox{for}~~ q=5   \\
  \end{array}
\right.~,
\end{eqnarray}
the (anti-)commutation relations in \eqref{QQM} are rewritten as 
\begin{eqnarray}
&&
\{\tilde Q^T,\tilde Q\}
=-4C\Gamma^\mu  p_- \tilde P_\mu~,
~~~
\{\tilde S^T,\tilde S\}
=-4C\Gamma^\mu  p_+\tilde K_\mu~,
\cr&&
\{\tilde Q^T,\tilde S\}
=\epsilon C\CI\Gamma^{\mu\nu} p_+ \tilde J_{\mu\nu}
-2 C\Gamma^{q+1}p_+\tilde D
-\beta C\Gamma^{a'}p_+  P_{a'}
-\epsilon\frac{\beta}{2}
C\CI\Gamma^{a'b'} p_+ J_{a'b'}~,
\cr&&
[\tilde P_\mu,\tilde S]=\epsilon\frac{1}{2}\tilde Q\Gamma_{\mu q+1}~,~~~
[\tilde K_\mu,\tilde Q]=-\epsilon\frac{1}{2}\tilde S\Gamma_{\mu q+1}~,~~~
[\tilde D, \tilde Q]=\frac{1}{2}\tilde Q ~,~~~
[\tilde D, \tilde S]=-\frac{1}{2}\tilde S
\cr&&
[\tilde J_{\mu\nu}, \tilde Q]=\frac{1}{2}\tilde Q\Gamma_{\mu\nu}~,~~
[\tilde J_{\mu\nu},\tilde S]=\frac{1}{2}\tilde S\Gamma_{\mu\nu}~,~~
[J_{a'b'},\tilde Q]=\frac{1}{2}\tilde Q\Gamma_{a'b'}~,~~
[J_{a'b'},\tilde S]=\frac{1}{2}\tilde S\Gamma_{a'b'}~,~~
\cr&&
[P_{a'},\tilde Q]=-\frac{1}{2}\tilde Q\CI\Gamma_{a'}~,~~~
[P_{a'},\tilde S]=-\frac{1}{2}\tilde S\CI\Gamma_{a'}~,
\label{Q conformal}
\end{eqnarray}
where the following relations have been used
\begin{eqnarray}
p_\pm^TC=\left\{
  \begin{array}{ll}
Cp_\pm       &~~\mbox{for}~~q=2    \\
Cp_\mp       &~~\mbox{for}~~q=5    \\
  \end{array}
\right.
.
\end{eqnarray}
The (anti-)commutation relations in \eqref{conformal}, \eqref{S} and \eqref{Q
conformal} form the ($q$+1)-dimensional superconformal algebra.  $\tilde
Q$ are 16 supercharges and $\tilde S$ are 16 superconformal charges.

\medskip 

Further decomposition of the generators as \eqref{decomposition 2}
reduces the (anti-)commutation relations in \eqref{Q conformal}  to
\begin{eqnarray}
\{\tilde Q^T,\tilde Q\}&=&
-4C\Gamma^+p_- P_+
-4C\Gamma^-p_- P_-
-4C\Gamma^ip_- P_i~,
\cr
\{\tilde S^T,\tilde S\}&=&
-4C\Gamma^+p_+ K_+
-4C\Gamma^-p_+ K_-
-4C\Gamma^ip_+ K_i~,
\cr
\{\tilde Q^T,\tilde S\}&=&
\epsilon C\CI\Gamma^{ij} p_+ J_{ij}
+2\epsilon C\CI\Gamma^{i+}p_+J_{i+}
+2\epsilon C\CI\Gamma^{i-} p_+ J_{i-}
\cr&&
+2C\Gamma^{q+1}\Gamma^{+}\Gamma^-p_+ D'
+2C\Gamma^{q+1}\Gamma^{-}\Gamma^+ p_+ D
-\beta C\Gamma^{a'}p_+P_{a'}
-\epsilon\frac{\beta}{2}C\CI\Gamma^{a'b'}p_+J_{a'b'}
~,
\cr
[K_\pm,\tilde Q]&=&
\epsilon \frac{1}{2}\tilde S\Gamma^{\mp}\Gamma_{q+1}~,~~~
[K_i,\tilde Q]=-\epsilon \frac{1}{2}\tilde S \Gamma_{iq+1} ~,~~~
\cr
[P_\pm,\tilde S]&=&-\epsilon \frac{1}{2}\tilde Q\Gamma^{\mp}\Gamma_{q+1}~,~~~
[P_i,\tilde S]=\epsilon \frac{1}{2}\tilde Q \Gamma_{iq+1} ~,~~~
\cr
[J_{ij},\tilde Q]&=&\frac{1}{2}\tilde Q \Gamma_{ij} ~,~~~
[J_{ij},\tilde S]=\frac{1}{2}\tilde S \Gamma_{ij} ~,~~~
[J_{i\pm},\tilde Q]=-\frac{1}{2}\tilde Q \Gamma_i\Gamma^\mp~,~~~
[J_{i\pm},\tilde S]=-\frac{1}{2}\tilde S \Gamma_i \Gamma^\mp~,
\cr
[D,\tilde Q]&=&-\frac{1}{4}\tilde  Q \Gamma^{+}\Gamma^-~,~~~
[D,\tilde S]=\frac{1}{4}\tilde  S \Gamma^-\Gamma^+~,~~~
\cr 
[D',\tilde Q]&=&-\frac{1}{4}\tilde  Q \Gamma^-\Gamma^+~,~~~
[D',\tilde S]=\frac{1}{4}\tilde  S \Gamma^+\Gamma^-~,~~~
\cr
[J_{a'b'},\tilde Q]&=&\frac{1}{2}\tilde Q\Gamma_{a'b'}~,~~~
[J_{a'b'},\tilde S]=\frac{1}{2}\tilde S\Gamma_{a'b'}~,~~~
\cr 
[P_{a'},\tilde Q]&=&-\frac{1}{2}\tilde Q\CI\Gamma_{a'}~,~~~
[P_{a'},\tilde S]=-\frac{1}{2}\tilde S\CI\Gamma_{a'}~,
\label{Q full}
\end{eqnarray}
where $\Gamma^\pm = \frac{1}{\sqrt{2}}(\Gamma^0\pm\Gamma^q)$ have been
introduced. 

\medskip 

As seen in the previous subsection,
$\{
P_\pm,~
P_i,~
K_+,~
J_{ij},~
J_{i+},~
D
\}
$
forms the Schr\"odinger algebra \eqref{Sch algebra}.
In the following, we shall show that
\begin{eqnarray}
\{
P_\pm,~
P_i,~
K_+,~
J_{ij},~
J_{i+},~
D,~
P_{a'},~
J_{a'b'},~
\tilde Q,~
S
\}
\label{super Sch generators}
\end{eqnarray}
is a subalgebra of osp(8$|$4) for $q=2$ or  osp(8$^*|$4) for $q=5$.
Here $S$ can be read off
from the following decomposition of $\tilde S$
\begin{eqnarray}
\tilde S= S +S' ~,~~~
S= \tilde S\ell_-~,~~~
S'= \tilde S \ell_+~,~~~
\ell_\pm=\frac{1}{2}(1\pm \Gamma^{0q})=
-\frac{1}{2}\Gamma^\pm\Gamma^\mp~.
\end{eqnarray}
Note that $\ell_\pm$ commute with $p_\pm$.
The superalgebra \eqref{super Sch generators} contains the Schr\"odinger
algebra and so($10-q$) as its bosonic subalgebra. Thus this superalgebra 
should also be referred to as an extended super Schr\"odinger algebra.

\medskip 

First derive the anti-commutation relations containing $\tilde Q$ and
$S$ only,
\begin{eqnarray}
\{\tilde Q^T,\tilde Q\}&=&
-4C\Gamma^+p_-P_+
-4C\Gamma^-p_-P_-
-4C\Gamma^ip_-P_i~,
\cr
\{S^T,S\}&=&-4C\Gamma^+\ell_-p_+ K_+~,
\cr
\{\tilde Q^T, S\}&=&
\epsilon C\CI\Gamma^{ij}\ell_-p_+ J_{ij}
+2\epsilon C\CI\Gamma^{i+}\ell_-p_+J_{i+}
+2C\Gamma^{q+1}\Gamma^{-}\Gamma^+ \ell_-p_+D
\cr&&
-\beta C\Gamma^{a'}\ell_-p_+ P_{a'}
-\epsilon \frac{\beta}{2}
C\CI\Gamma^{a'b'}\ell_- p_+ J_{a'b'}
~.
\end{eqnarray}
The generators appearing in the right-hand sides are contained in
\eqref{super Sch generators}. Next we derive commutation relations
between bosonic generators and fermionic generators in \eqref{super Sch
generators}
\begin{eqnarray}
&&
[K_+,\tilde Q]=\epsilon \frac{1}{2}S\Gamma^{-}\Gamma_{q+1}~,~~~
[P_-,S]=-\epsilon \frac{1}{2}\tilde Q\ell_+\Gamma^{+}\Gamma_{q+1}~,~~~
[P_i, S]=\epsilon \frac{1}{2}\tilde Q\ell_- \Gamma_{i{q+1}} ~,~~~
\cr&&
[J_{ij},\tilde Q]=\frac{1}{2}\tilde Q \Gamma_{ij} ~,~~~
[J_{ij}, S]=\frac{1}{2} S \Gamma_{ij} ~,~~~
[J_{i+},\tilde Q]=-\frac{1}{2}\tilde Q \Gamma_i\Gamma^-~,~~~
\cr&&
[D,\tilde Q]=-\frac{1}{4}\tilde  Q \Gamma^{+}\Gamma^-~,~~~
[D,  S]=\frac{1}{4}   S \Gamma^-\Gamma^+~,~~~
\cr&&
[J_{a'b'},\tilde Q]=\frac{1}{2}\tilde Q\Gamma_{a'b'}~,~~~
[J_{a'b'}, S]=\frac{1}{2} S\Gamma_{a'b'}~,~~~
\cr&&
[P_{a'},\tilde Q]=-\frac{1}{2}\tilde Q\CI\Gamma_{a'}~,~~~
[P_{a'}, S]=-\frac{1}{2} S\CI\Gamma_{a'}~. \label{MQS}
\end{eqnarray}
Here $\tilde{Q}$ and $S$ only appear in the right-hand sides of the
commutators in (\ref{MQS}). Thus we have found an extended super
Schr\"odinger subalgebra of osp(8$|$4) and osp(8$^*|$4). The number of
the remaining supercharges is 24 since 1/4 supercharges have been
projected out.

\medskip 

Note that the extended super Schr\"odinger subalgebra further contains a
smaller super subalgebra formed by the set of the generators,
\begin{eqnarray}
\{
P_\pm,~
P_i,~
K_+,~
J_{ij},~
J_{i+},~
D,~
P_{a'},~
J_{a'b'},~
Q
\}\,, 
\label{Sch reduced}
\end{eqnarray}
where the generator $Q$ is obtained via a projection, 
\[
\tilde Q= Q +Q' \,,~~~
Q= \tilde Q \ell_-\,~~~
Q'= \tilde Q\ell_+\,.
\]  
The commutation relations including $Q$ are 
\begin{eqnarray}
&&
\{Q^T,Q\}=-4C\Gamma^+ \ell_- p_- P_+~,~~~
[J_{ij},Q]=\frac{1}{2}Q\Gamma_{ij}~,~~~
\cr&&
[J_{a'b'}, Q]=\frac{1}{2} Q\Gamma_{a'b'}~,~~~
[P_{a'}, Q]=-\frac{1}{2} Q\CI\Gamma_{a'}~.
\end{eqnarray}
The superalgebra \eqref{Sch reduced} can further be reduced to a super
subalgebra
\begin{eqnarray}
\{P_\pm,~P_i,~ K_+,~ J_{ij},~ J_{i+},~ D,~ Q
\}~.
\end{eqnarray}
This is a super Schr\"odinger algebra with eight supercharges. Its bosonic
subalgebra contains the original Schr\"odinger algebra only.

\medskip

The interpretation of so($10-q$) is similar to the case of so(6)
explained in subsection 2.3.  The so(8) for $q=2$ acts on eight
two-component spinors $\tilde Q$ and eight two-component spinors $\tilde
S$ in $(2+1)$-dimensions, separately.  In $(1+1)$-dimensions, as $p_\pm$
is the chirality projector, eight two-component spinors reduce to a pair
of eight one-component spinors.  The so(8) acts on four of eight
one-component spinors separately.  On the other hand, the
so(5) $\cong$ sp(4) for $q=5$ acts on a pair of four four-component Weyl spinors 
in ($5+1$)-dimensions where we note
$p_\pm$ is the chirality projector in ($5+1$)-dimensions.  
The sp(4) rotation imposes a symplectic Majorana condition on the spinors,
and we are left with 
a pair of four four-component symplectic Majorana-Weyl spinors. 
The reduction
to $(4+1)$-dimensions reduces them to four of four four-component
spinors subject to the sp(4) rotation.  
Thus we are left with four of two four-component symplectic Majorana
 spinors in $(4+1)$-dimensions.
The so(5) $\cong$ sp(4) acts on the four sets of spinors, separately.  
Thus so($10-q$) acts as R-symmetry even in the
extended super Schr\"odinger algebra.

\section{Conclusion and Discussion}

We have found (extended) super \sh algebras contained in the
psu(2,2$|$4). An extended super \sh subalgebra contains, as well as
generators of the original \sh algebra, 24 supercharges (16 rigid
supersymmetries and half of 16 superconformal) and generators of
so(6). It also contains a smaller subalgebra. It is composed of
generators of the original \sh algebra, eight supercharges (half of 16
rigid supersymmetries) and the generators of so(6). It is still a
subalgebra even if there are no so(6) generators. 
We have also discussed super \sh subalgebras
of the superconformal algebras, osp(8$|$4) and osp(8$^{\ast}|$4). The
results are similar to the case of psu(2,2$|$4).

\medskip 

One may find another super \sh subalgebra other than the ones found
here.  It is a nice subject to completely classify (extended) super \sh
subalgebras of the superconformal algebras. It would also be interesting
to find super \sh subalgebras of the other superconformal algebras besides
psu(2,2$|4$), osp(8$|$4) and osp(8$^{\ast}|$4).

\medskip 

The next issue is to consider the coset construction by using the
supergroup of the super \sh algebra obtained here. In the usual coset
construction the isometry group is divided by the local Lorentz
symmetry, as is well known for the AdS$_5\times$S$^5$\,. The local
Lorentz symmetry should be replaced by something different in the case
of the \sh group.  By finding an appropriate coset, it might be possible
to reproduce the asymptotically plane-wave background found in \cite{Son,BM}. 
Our super
\sh algebra directly appeared from the psu(2,2$|4$) and hence it may
lead to a supersymmetric
background
constructed
from the AdS$_5\times$S$^5$  somehow. 

\medskip

It would also be important to construct the CFT action, which has the super \sh symmetry obtained here
as the maximal symmetry. 
It would be a clue to
shed light on the non-relativistic holographic relation.

\section*{Acknowledgment}

The authors would like to thank H.Y.~Chen, S.~Ryu and A.~Schnyder for
useful discussion. The work of M.S.\ was supported in part by the
Grant-in-Aid for Scientific Research (19540324) from the Ministry of
Education, Science and Culture, Japan. The work of K.Y.\ was supported
in part by JSPS Postdoctoral Fellowships for Research Abroad and the
National Science Foundation under Grant No.\,NSF PHY05-51164.
K.~Y. also thanks the workshop ``Physics of the Large Hadron Collider''
at KITP. Discussions with the participants gave him an opportunity to
consider the issue of this paper.

\end{document}